\documentclass[10pt,final,a4paper]{iopart}
\usepackage{epsfig}
\usepackage{subfigure}

\begin{document}

\title[Computational studies of light in straight and curved
	fibres]{Computational studies of light acceptance and propagation 
	in straight and curved multimodal active fibres} 
\author{C P Achenbach\footnote{
    Present address: Institut f{\"u}r 
    Kernphysik, Joh.\ Gutenberg-Universit{\"a}t Mainz, 
    J\,J\,Becher-Weg 45, 55099 Mainz, Germany.} 
  and J H Cobb}
\address{University of Oxford, Sub-department of Particle Physics, 
	Denys Wilkinson Building, Keble Road, Oxford OX1 3RH, UK}
\ead{p.achenbach@physics.ox.ac.uk}

\begin{abstract}
    A Monte Carlo simulation has been performed to track light rays in
    cylindrical multimode fibres by ray optics. The trapping efficiencies
    for skew and meridional rays in active fibres and distributions of
    characteristic quantities for all trapped light rays have been
    calculated. The simulation provides new results for curved fibres,
    where the analytical expressions are too complex to be solved. The
    light losses due to sharp bending of fibres are presented as a
    function of the ratio of curvature to fibre radius and bending
    angle. It is shown that a radius of curvature to fibre radius ratio of
    greater than 65 results in a light loss of less than 10\% with the
    loss occurring in a transition region at bending angles
    $\Phi \sim \pi/8$\,rad.\\

\end{abstract}

\jl{11}
\submitted

\section{Introduction}
Active optical fibres are becoming more and more important in the
field of detection and measurement of ionising radiation and
particles. Light is generated inside the fibre either through
interaction with the incident radiation (scintillating fibres) or
through absorption of primary light (wavelength-shifting
fibres). Plastic fibres with large core diameters, i.e.\ where the
wavelength of the light being transmitted is much smaller than the
fibre diameter, are commercially available and readily fabricated,
have good timing properties and allow a multitude of different
geometrical designs. The low costs of plastic materials make it
possible for many present day or future experiments to use such fibres
in large quantities (see~\cite{Leutz1995} for a review of
active fibres in high energy physics). For the construction of the
highly segmented tracking detector of the ATLAS experiment approved
for the LHC collider at CERN more than 600,000 wavelength-shifting
fibres have been used~\cite{ATLAS1994}. Our work is also motivated by
the fact that spiral fibres embedded in scintillators are being used
for calorimetric measurements in long base line neutrino oscillation
experiments, most recently in the MINOS experiment~\cite{MINOS1998}.

The treatment of small diameter optical fibres involves
electromagnetic theory applied to dielectric waveguides, which was
first achieved by Snitzer~\cite{Snitzer1961} and Kapany {\it et
al}~\cite{Kapany1963}. Although this approach provides insight into
the phenomenon of total internal reflection and eventually leads to
results for the field distributions and electromagnetic radiation for
curved fibres, it is advantageous to use ray optics for applications
to large diameter fibres where the waveguide analysis is an
unnecessary complication. In ray optics a light ray may be categorised
by its path along the fibre. The path of a meridional ray is confined
to a single plane, all other modes of propagation are known as skew
rays. The optics of meridional rays in fibres was developed in the
1950s~\cite{Kapany1957} and can be found in numerous textbooks, e.g.\
in ~\cite{Kapany1967,Allan1973,Ghatak1998}. Since then, the scientific
and technological progress in the field of fibre optics has been
enormous. Despite the extensive coverage of theory and experiment in
this field, only fragmentary studies on the trapping efficiencies and
refraction of skew rays in curved multimode fibres could be
found~\cite{Winkler1979,Badar1989,Badar1991A,Badar1991B}. We have
therefore performed a three-dimensional simulation of photons
propagating in simple circularly curved fibres in order to quantify
the losses and to establish the dependence of these losses on the
angle of the bend. We have also briefly investigated the time
dispersion in fibres. For our calculations a common type of fibre in
particle physics is assumed, specified by a polystyrene core of
refractive index $n_{\it core}=$ 1.6 and a thin polymethylmethacrylate
(PMMA) cladding of refractive index $n_{\it clad}=$ 1.49, where the
indices are given at a wavelength of 590\,nm. Another common cladding
material is fluorinated polymethacrylate with $n_{\it clad}=
1.42$. Typical diameters are in the range of 0.5 -- 1.5\,mm.

This paper is organised as follows: section~2 describes the analytical
expressions of trapping efficiencies for skew and meridional rays in
active, i.e.\ light generating, fibres. The analytical description of
skew rays is too complex to be solved for sharply curved fibres and
the necessity of a simulation becomes evident. In section~3 a
simulation code is outlined that tracks light rays in cylindrical
fibres governed by a set of geometrical rules derived from the laws of
optics. section~4 presents the results of the simulations. These
include distributions of the characteristic properties which describe
light rays in straight and curved fibres, where special emphasis is
placed on light losses due to the sharp bending of fibres. Light
dispersion is briefly reviewed in the light of the results of the
simulation. The last section provides a short summary.

\section{Trapping of photons}
When using scintillating or wavelength-shifting fibres in charged
particle detectors the trapped light as a fraction of the intensity of
the emitted light is very important in determining the light yield of
the system. For very low light intensities as encountered in many
particle detectors the photon representation is more appropriate to
use than a description by light rays. Whether the fibres are
scintillating or wavelength-shifting one is only ever concerned with a
few 10's or 100's of photons propagating in the fibre and single
photon counting is often necessary.

The geometrical path of any rays in optical fibres, including skew
rays, was first analysed in a series of papers by
Potter~\cite{Potter1961} and Kapany~\cite{Kapany1961}. The treatment
of angular dependencies in our paper is based on that. The angle
$\gamma$ is defined as the angle of the projection of the light ray in
a plane perpendicular to the axis of the fibre with respect to the
normal at the point of reflection. One may describe $\gamma$ as a
measure of the `skewness' of a particular ray, since meridional rays
have this angle equal to zero. The polar angle, $\theta^\prime$, is
defined as the angle of the light ray in a plane containing the fibre
axis and the point of reflection with respect to the normal at the
point of reflection. It can be shown that the angle of incidence at
the walls of the cylinder, $\alpha$, is given by $\cos{\alpha}=
\cos{\theta^\prime}\, \cos{\gamma}$. The values of the two orthogonal
angles $\theta^\prime$ and $\gamma$ will be preserved independently
for a particular photon at every reflection along its path.

In general for any ray to be internally reflected within the cylinder
of the fibre, the inequality $\sin{\alpha} \geq
\sin{\theta^\prime_{\it crit}} = n_{\it clad}/n_{\it core}$ must be
fulfilled, where the critical angle, $\theta^\prime_{\it crit}$, is
given by the index of refraction of the fibre core, $n_{\it core}$,
and that of the cladding, $n_{\it clad}$. In the meridional
approximation the above equations lead to the well known critical
angle condition for the polar angle, $\theta^\prime \ge
\theta^\prime_{\it crit}$, which describes an acceptance cone of
semi-angle, $\theta\ [= \pi/2 - \theta^\prime]$, with respect to the
fibre axis (see for example~\cite{Potter1961} and references
therein). Thus, in this approximation all light within the forward
cone will be considered as trapped and undergo multiple total internal
reflections to emerge at the end of the fibre.

\begin{figure}[htbp]
  \begin{center}
    \subfigure[]{
	\epsfig{width= 0.47 \textwidth, file= 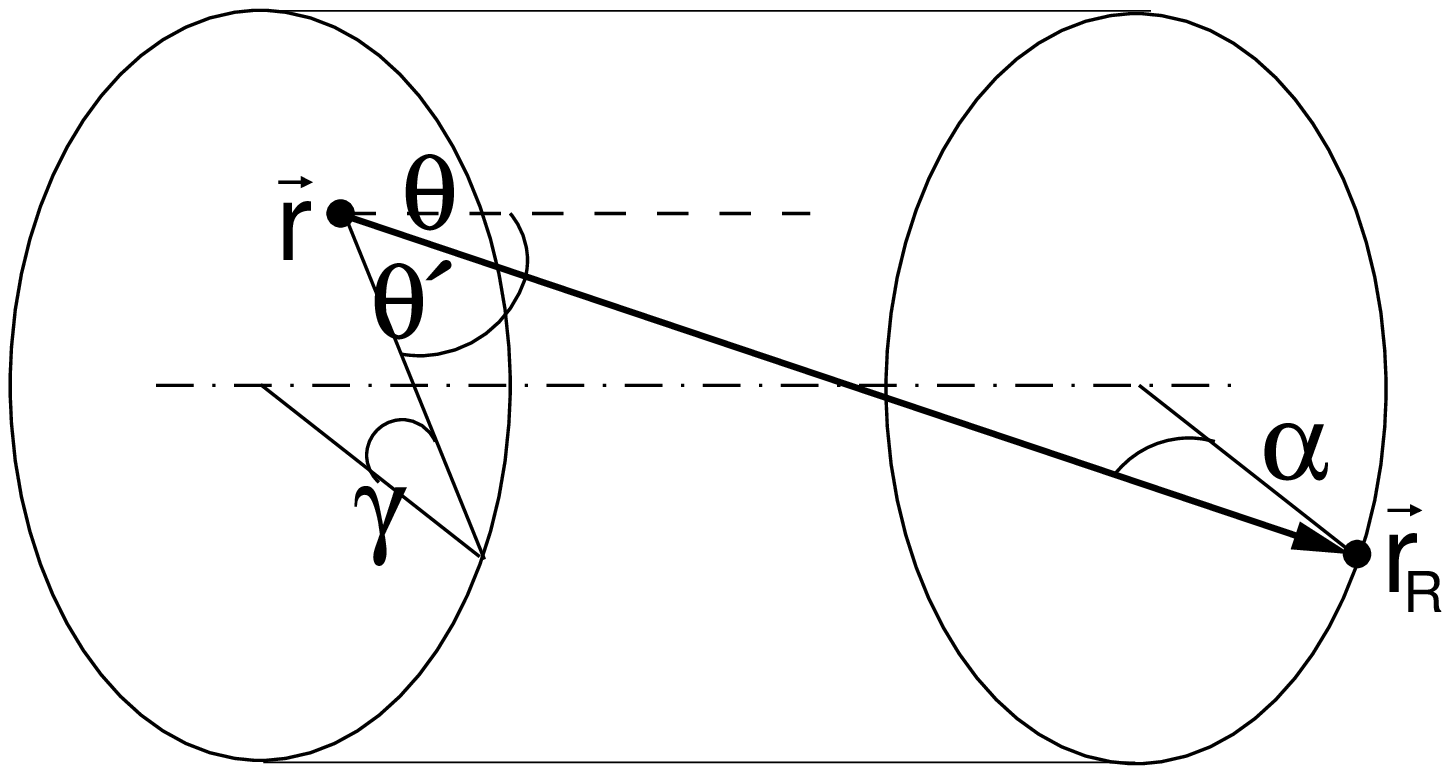} }
    \subfigure[]{
	\epsfig{width= 0.47 \textwidth, file= 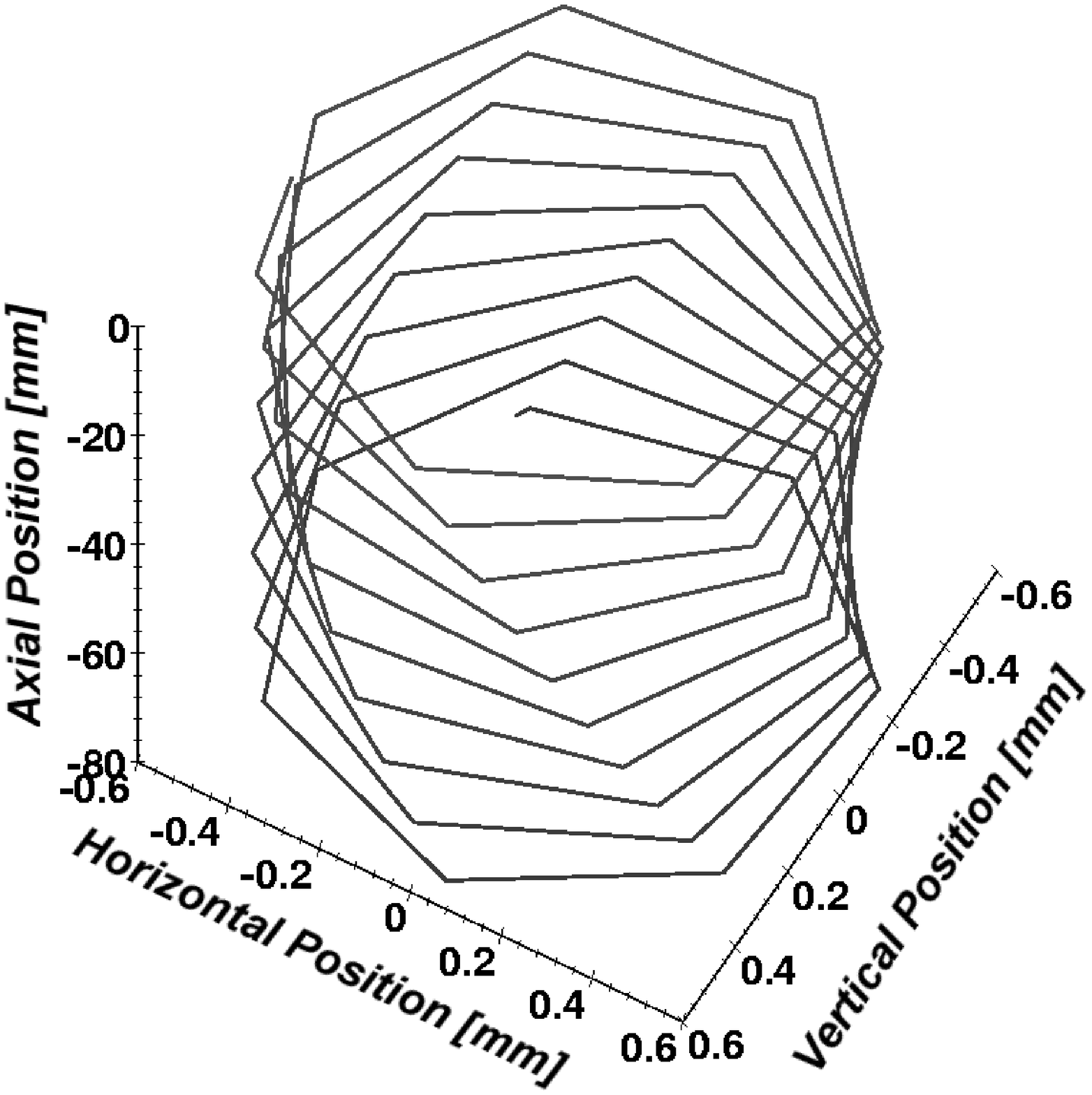} } 
    \caption{ (a) Definition of the angles for a skew ray from an 
	initial point, $\vec{r}$ to a point of reflection, $\vec{r}_R$,
	on the core-cladding interface. The axial angle of the ray is 
	$\theta$. The angle between the projection of the ray in a plane 
	perpendicular to the axis of the fibre with respect to the normal 
	at the point of reflection, $\gamma$, defines the ray's 
	`skewness'. The angle of incidence at the point of reflection 
	is $\alpha$. 
	(b) The helical path of a skew ray in a fibre of radius 
	$\rho = 0.6\,$mm which spirals along the core-cladding 
	interface. The $z$-axis represents the axial fibre 
	length and the fibre cross-section lies in the $xy$-plane.}
    \label{fig:description}
  \end{center}
\end{figure}

For the further discussion in this paper it is convenient to use the
axial angle, $\theta$, as given by the supplement of $\theta^\prime$,
and the skew angle, $\gamma$, to characterise any light ray in terms
of its orientation, see figure~\ref{fig:description} for an
illustration.

The flux transmitted by a fibre is determined by an integration over
the angular distribution of the light emitted within the acceptance
domain, i.e.\ the phase space of possible propagation modes. Using the
expression given by Potter {\it et al}~\cite{Potter1963} and setting the
transmission function, which parameterises the light attenuation, to
unity the light flux can be written as follows:
\begin{equation}
\eqalign{
  F & =  F_m + F_s\\
    & =  4 \rho^2 \int_{\theta= 0}^{\theta_{\it crit}} 
	\int_{\gamma= 0}^{\pi/2} \int_{\phi= 0}^{\pi/2} 
	I(\theta,\phi)\, \cos^2{\gamma}\, d\gamma\, d\Omega\ +\\
    & 4 \rho^2 \int_{\theta= \theta_{\it crit}}^{\pi/2} 
	\int_{\gamma= \overline{\gamma}(\theta)}^{\pi/2} 
        \int_{\phi= 0}^{\pi/2}
	I(\theta,\phi)\, \cos^2{\gamma}\, d\gamma\, d\Omega\ ,
}
\end{equation}
where $d\Omega$ is the element of solid angle,
$\overline{\gamma}(\theta)$ refers to the maximum axial angle allowed
by the critical angle condition, $\rho$ is the radius of a cylindrical
fibre and $I(\theta,\phi)$ is the angular distribution of the emitted
light in the fibre core. The two terms, $F_m$ and $F_s$, refer to
either the meridional or skew cases, respectively. The lower limit of
the integral for $F_s$ is $\overline{\gamma}=
\arccos{(\sin{\theta_{\it crit}}/\sin{\theta})}$.

The trapping efficiency for forward propagating photons,
$\epsilon^{1/2}$, may be defined as the fraction of totally internally
reflected photons. The formal expression for the trapping efficiency,
including skew rays, is derived by dividing the transmitted flux by
the total flux through the cross-section of the fibre core, $F_0$.
For isotropic emission of fluorescence light the total flux equals $4
\pi^2 \rho^2 I_0$. Then, the first term of equation~(1) gives the
trapping efficiency in the meridional approximation,
\begin{equation}
  \epsilon^{1/2}_m = F_m/F_0 = \frac{1}{2} (1 - 
	\cos{\theta_{\it crit}}) \approx 
	\frac{\theta^2_{\it crit}}{4}\ ,
  \label{eq:omega_m}
\end{equation}
where all photons are considered to be trapped if $\theta \le
\theta_{\it crit}$, independent of their actual skew angles. 

The integration of the second term of equation~(1) gives the
contributions of all skew rays to the trapping efficiency. Integrating
by parts, one obtains
\begin{equation}
  \hspace{-1cm} \epsilon^{1/2}_s = \frac{1}{2} \cos{\theta_{\it crit}} -
  \frac{\cos^2{\theta_{\it crit}} \sin{\theta_{\it crit}}}{2\pi}
  \int_0^1 \frac{dt}{\sqrt{(1-t)\,t}\, \left(1-t \cos^2{\theta_{\it
  crit}} \right)}\ ,
\end{equation}
with $t= \cos^2{\theta}/\cos^2{\theta_{\it crit}}$. Complex
integration leads to the result:
\begin{equation}
  \epsilon^{1/2}_s = \frac{1}{2} (1 - \cos{\theta_{\it crit}})
  \cos{\theta_{\it crit}}\ .
  \label{eq:omega_s}
\end{equation}
The total initial trapping efficiency is then: 
\begin{equation}
  \epsilon^{1/2} = \frac{1}{2} (1 - \cos^2{\theta_{\it crit}}) 
	\approx \frac{\theta^2_{\it crit}}{2}\ ,
  \label{eq:omega_tot}
\end{equation}
which is approximately twice the trapping efficiency in the meridional
approximation for small critical angles. The trapping efficiency of
rays is crucially dependent on the circular symmetry of the
core-cladding interface. Any ellipticity or variation in the fibre
diameter will lead to the refraction of some skew rays. Furthermore,
skew rays have a much longer optical path length, suffer from more
reflections and therefore get attenuated more quickly, see section~4
for a quantitative analysis of this effect. In conclusion, for long
fibres the effective trapping efficiency is closer to
$\epsilon^{1/2}_m$ than to $\epsilon^{1/2}$. Formula~\ref{eq:omega_m}
yields a trapping efficiency of $\epsilon^{1/2}_m=$ 3.44\% for plastic
fibres with $n_{\it core}=$ 1.6 and $n_{\it clad}=$ 1.49. For these
``standard'' parameters the efficiency in formula~\ref{eq:omega_s}
evaluates to $\epsilon^{1/2}_s=$ 3.20\% and in
formula~\ref{eq:omega_tot} to $\epsilon^{1/2}=$ 6.64\%.

\section{Description of the tracking code}
The aim of the program is to track light rays through a fibred. Since
the analytic analysis of the passage of skew rays along a curved fibre
is exceedingly complex we treat the problem with a Monte Carlo
technique. This type of numerical integration using random numbers is
a standard method in the field of particle physics and is now
practical given the CPU power currently available. On its path the ray
is subject to attenuation, parameterised firstly by an effective
absorption coefficient and secondly by a reflection coefficient. At
the core-cladding interface the ray can be reflected totally or
partially internally. In the latter case a random number is compared
to the reflection probability to select reflected rays.

Light rays are randomly generated on the circular cross-section of a
fibre with radius $\rho$. An arbitrary ray is defined by its axial and
azimuthal or skew angle. An advantage of this method is that any
distribution of light rays can easily be generated. The axis of the
fibre is defined by a curve $z= f(s)$ where $s$ is the arc length. For
$s < 0$, it is a straight fibre along the negative $z$-axis and for $0
< s < L_F$, the fibre is curved in the $xz$-plane with a radius of
curvature $R_{\it curve}$. In particular, the curve $f(s)$ is
tangential to the $z$-axis at $s = 0$.

Light rays are represented as lines and determined by two points,
$\vec{r}$ and $\vec{r}^{\,\prime}$. The points of incidence of rays
with the core-cladding interface are determined by solving the
appropriate systems of algebraic equations. In the case of a straight
fibre the geometrical representation of a straight cylinder is used
resulting in the quadratic equation
\begin{equation} 
  (x + (x^\prime - x)\times m)^2 + (y +
  (y^\prime - y)\times m)^2 - \rho^2 = 0\ .
\end{equation}
The positive solution for the parameter $m$ defines the point of
incidence, $\vec{r}_R$, on the cylinder wall. In the case of a fibre
curved in a circular path, the cylinder equation is generalized by the
torus equation
\begin{equation}
  \hspace{-1cm} \eqalign{ ( R_{\it curve} - ( (x + (x^\prime - 
	x)\times m + R_{\it curve})^2\\
	+ (z + (z^\prime - z)\times m)^2 )^{1/2} )^2\\
  	+ (y + (y^\prime - y)\times m)^2 - \rho^2 = 0\ .}
\end{equation} 
The coefficients of this fourth degree polynomial are real and depend
only on $R_{\it curve}$ and the vector components of $\vec{r}$ and
$\vec{r}^{\,\prime}$ up to the fourth power. In most cases there are
two real roots, one for the core-cladding intersection in the forward
direction and one at $\vec{r}$ if the initial point already lies on
the cylinder wall. The roots are found using Laguerre's
method~\cite{Recipes1992}. It requires complex arithmetic, even while
converging to real roots, and an estimate for the root to be
found. The routine implements a stopping criterion in case of
non-convergence because of round-off errors. The initial estimate is
given by the intersection point of the light ray and a straight
cylinder that has been rotated and translated to the previous
reflection point. A driver routine is used to apply Laguerre's method
to all four roots and to perform the deflation of the remaining
polynomial. Finally the roots are sorted by their real part. The
smallest positive, real solution for $m$ is then used to determine the
reflection point, $\vec{r}_R$.

After the point of incidence has been found, the reflection length and
absorption probability can be calculated. The angle of incidence,
$\alpha$, is given by $\cos{\alpha} = \vec{r}_{in} \cdot \vec{n}$,
where $\vec{n}$ denotes the unit vector normal to the core-cladding
interface at the point of reflection and $\vec{r}_{in}=
(\vec{r}-\vec{r}_R)/|\vec{r}-\vec{r}_R|$ is the unit incident
propagation vector. Now the reflection probability corresponding to
this angle $\alpha$ is determined. In case the ray is partially or
totally internally reflected the total number of reflections is
increased and the unit propagation vector after reflection,
$\vec{r}_{\it out}$, is then calculated by mirroring $\vec{r}_{in}$
with respect to the normal vector: $\vec{r}_{\it out} = \vec{r}_{in} -
2 \vec{n} \cos{\alpha}$. The program returns in a loop to the
calculation of the next reflection point. When the ray is absorbed on
its path or not reflected at the reflection point the next ray is
generated at the fibre entrance end. A scheme on the main steps of the
program can be found in figure~\ref{fig:scheme}. At any point of the
ray's path its axial, azimuthal and skew angle are given by scalar
products of the ray vector with the coordinate axes in a projection on
a plane perpendicular to the fibre axis and parallel to the fibre
axis, respectively. The transmitted flux of a specific fibre, taking
all losses caused by bending, absorption and reflections into account,
is calculated from the number of lost rays compared to the number of
rays reaching the fibre exit end.

This method gives rise to an efficient simulation technique for fibres
with constant curvature. It is possible to extend the method for the
study of arbitrarily curved fibres by using small segments of constant
curvature. In the current version of the program light rays are
tracked in the fibre core only and no tracking takes place in the
surrounding cladding, corresponding to infinite cladding thickness. In
long fibres cladding modes will eventually be lost, but for lengths $<
1$\,m they can contribute to the transmission function. The
simulation code is written in Fortran and it takes about 1.5\,ms to
track a skew ray through a curved fibre.

\section{Results of the tracking code}
Figure~\ref{fig:description} shows the passage of a skew ray along a
straight fibre. The light ray has been generated off-axis with an
axial angle of $\theta= 0.42$ and would not be trapped if it were
meridional.  In general, the projection of a meridional ray on a plane
perpendicular to the fibre axis is a straight line, whereas the
projection of a skew ray changes its orientation with every
reflection. In the special case of a cylindrical fibre all meridional
rays pass through the fibre axis. The figure illustrates the
preservation of the skew angle, $\gamma$, during the propagation.

\subsection{Trapping efficiency and acceptance domain}
Figure~\ref{fig:phasespace}(a) shows the total acceptance domain and
its splitting into the meridional and skew regions in the meridional
ray approximation. The phase space density, i.e.\ the number of
trapped rays per angular interval, is represented by proportional
boxes. The density increases with $\cos^2{\gamma}$ and $\sin{\theta}$.
The contours relate to sharply curved fibres and are explained in
section~\ref{section:curvedfibres}. Figure~\ref{fig:phasespace}(b)
shows a projection of the phase space onto the $\sin\theta$-axis. A
peak around the value $\sin{\theta_{\it crit}}$ is apparent. A skew
ray can be totally internally reflected at larger angles $\theta$ than
meridional rays and the relationship between the minimum permitted
skew angle, $\overline{\gamma}$, at a given axial angle, $\theta$, is
determined by the critical angle condition: $\cos{\overline{\gamma}}=
\sin{\theta_{\it crit}} / \sin{\theta}$.

Photons are generated randomly on the cross-section of the fibre with
an isotropic angular distribution in the forward direction. The figure
gives values for the two trapping efficiencies which can be determined
by integrating over the two angular regions. The integrals are
identical in value to the expressions in formulae~\ref{eq:omega_m}
and~\ref{eq:omega_s}. It is obvious from the critical angle condition
that a photon emitted close to the cladding has a higher probability
to be trapped than when emitted close to the centre of the fibre. For
a given axial angle the range of possible azimuthal angles, in which
the photons are uniformly distributed, for the photon to get trapped
increases with the radial position, $\hat{\rho}$, of the light emitter
in the fibre core. It can be deduced from figure~\ref{fig:trap-r} that
the meridional approximation is a good estimate for $\epsilon$ if the
photons originate at radial positions $\hat{\rho} < 0.8$. The trapping
of skew rays only becomes significant for photons originating at
radial positions $\hat{\rho} \ge 0.9$. This fact has been discussed
before, e.g.\ in~\cite{Johnson1994}. Figure~\ref{fig:trap-theta} shows
the the trapping efficiency as a function of the axial angle. All
photons with axial angles below $\theta_{\it crit}$ are trapped in the
fibre, whereas photons with larger angles are trapped only if their
skew angle exceeds the minimum permitted skew angle. It can be seen
that the trapping efficiency falls off very steeply with the axial
angle.

\subsection{Propagation of photons}
The analysis of trapped photons is based on the total photon path
length per axial fibre length, $P$, the number of internal reflections
per axial fibre length, $\eta$, and the optical path length between
successive internal reflections, $l_R$, where we follow the
nomenclature of Potter and Kapany. It should be noted that these three
variables are not independent as $P= \eta \times l_R$.

Figure~\ref{fig:pathlength} shows the distribution of the normalised
path length, $P(\theta)$, for photons reaching the exit end of
straight and curved fibres of 0.6\,mm radius. The figure also gives
results for curved fibres of two different radii of curvature. The
distribution of path lengths shorter than the path length for
meridional photons propagating at the critical angle is almost
flat. It can easily be shown that the normalised path length along a
straight fibre is given by the secant of the axial angle and is
independent of other fibre dimensions: $P(\theta)= \sec\theta$. In
case of the curved fibre the normalised path length of the trapped
photons is less than the secant of the axial angle and photons on near
meridional paths are refracted out of the fibre most.

The distribution of the normalised number of reflections,
$\eta(\theta)$, for photons reaching the exit end of straight and
curved fibres is shown in figure~\ref{fig:reflections}. Again, the
figure gives results for curved fibres of two different radii of
curvature. The number of reflections a photon experiences scales with
the reciprocal of the fibre radius. In the meridional approximation
the normalised number of reflections is related by simple trigonometry
to the axial angle and the fibre radius: $\eta_m(\theta) =
\tan{\theta}/2\rho$. The distribution of $\eta_m$, based on the
distribution of axial angles for the trapped photons, is represented
by the dashed line. The upper limit, $\eta(\theta_{\it crit})$, is
indicated in the plot by a vertical line. The number of reflections
made by a skew ray, $\eta_s(\theta)$, can be calculated for a given
skew angle: $\eta_s(\theta)= \eta_m(\theta) / \cos{\gamma}$. It is
clear that this number increases significantly if the skew angle
increases. From the distributions it can be seen that in curved fibres
the trapped photons experience fewer reflections on average.

Figure~\ref{fig:rlambda}(a) shows the distribution of the reflection
length, $l_R(\theta)$, for photons reaching the exit end of fibres of
radius $\rho= 0.6$\,mm. The reflection length will scale with the
fibre radius. The left figure shows $l_R(\theta)$ for four different
over-all fibre lengths, $L_F=$ 0.5, 1, 2 and 3\,m, and the attenuation
characteristics of the fibre is made apparent by the non-vanishing
attenuation parameters used. Short reflection lengths correspond to
long optical path lengths and large numbers of reflections. Because of
the many reflections and the long total paths traversed, these photons
will be attenuated faster than photons with larger reflection
lengths. This reveals the high attenuation of rays with large skew
angles. In the meridional approximation the reflection length is
related to the axial angle by: $l_R= 2\rho/\cos{\theta}$. In the
figure the minimum reflection length allowed by the critical angle
condition is shown by a vertical line at $l_R(\theta_{\it crit})=
3.29$\,mm. On average photons propagate with smaller reflection
lengths along the curved fibre. Figure~\ref{fig:rlambda}(b) shows the
distribution of $l_R(\theta)$ in curved fibres of two different radii
of curvature, $R_{\it curve}=$ 2 and 8\,cm.  The fibre radius is
identical to the one used in figure (a) and the over-all fibre length
is 0.5\,m. It can be seen that the sharp peak in the photon
distribution flattens with decreasing radius of curvature, so that the
region of highest attenuation is close to the reflection length for
photons propagating at the critical angle.

In contrast to the analysis of straight fibres an approximation of the
sharply curved fibre by meridional rays is not a very good one, since
only a very small fraction of the light rays have paths lying in the
bending plane. It is clear that when a fibre is curved the path
length, the number of reflections and the reflection length of a
particular ray in the fibre are affected, which is clearly seen in
figures~\ref{fig:pathlength},~\ref{fig:reflections} and
\ref{fig:rlambda}(b). The over-all fibre length for the curved fibres
in these calculations is 0.5\,m and the fibres are curved for their
entire length. The average optical path length and the average number
of reflections in a fibre curved over a circular arc are less than
those for the same ray in a straight fibre for those photons which
remain trapped.

\subsection{Light attenuation}
Light attenuation in active fibres has many sources, among them
self-ab\-sorp\-tion, optical non-uni\-formities, reflection losses and
absorption by impurities. The two main sources of attenuation in this
type of fibres are the absorption of scintillation light and Rayleigh
scattering from small density fluctuations. The self-absorption is due
to the overlap of the emission and absorption bands of the fluorescent
dyes. The cumulative effect of these attenuation processes can be
conveniently parameterised by an effective attenuation length over
which the signal amplitude is attenuated to 1$/e$ of its original
value, a method often applied in high energy physics applications. The
attenuation of active fibres at wavelengths close to its emission band
($400-600$\,nm) is much higher than in wavelength regions of interest
for standard applications of communication fibres where mainly
infrared light is transmitted ($0.8-0.9\,\mu$m and $1.2-1.5\,\mu$m).

Restricting the analysis to these processes, the transmission through
an active fibre can be represented for any given axial angle by $T=
\exp\left[- P(\theta) L_F/\Lambda_{\it bulk}\right]\, \times
q^{\eta(\theta) L_F}$, where the exponential function describes light
losses due to bulk absorption and scattering (bulk absorption length
$\Lambda_{\it bulk}$), and the second factor describes light losses
due to imperfect reflections (reflection coefficient $q$) which can be
caused by a rough surface or variations in the refractive indices.  A
comparison of some of our own measurements to determine the
attenuation length of plastic fibres with other available data
indicates that a reasonable value for the bulk absorption length is
$\Lambda_{\it bulk} \sim 3$\,m. Most published data suggest a
deviation of the reflection coefficient, which parameterises the
internal reflectivity, from unity between $5 \times 10^{-5}$ and $6.5
\times 10^{-5}$ \cite{Ambrosio1991}. A reasonable value of $q= 0.9999$
is used in the simulation to account for all losses proportional to
the number of reflections.

Internal reflections being less than total give rise to so-called
leaky or non-guided modes, where part of the electromagnetic energy is
radiated away. Rays in these modes populate a region defined by axial
angles above the critical angle and skew angles slightly larger than
the ones for totally internally reflected photons. These modes are
taken into account by using the Fresnel equation for the reflection
coefficient, $\langle R \rangle$, averaged over the parallel and
orthogonal plane of polarisation
\begin{equation}
  \langle R \rangle = \frac{1}{2} \left( R_{||} + R_\perp \right) = 
     \frac{1}{2} \left( \frac{\tan^2(\alpha - \beta)}
     {\tan^2(\alpha + \beta)} + \frac{\sin^2(\alpha - \beta)}
     {\sin^2(\alpha + \beta)} \right)\ ,
\end{equation}
where $\alpha$ is the angle of incidence and $\beta$ is the refraction
angle. However, it is obvious that non-guided modes are lost quickly
in a small fibre. This is best seen in the fraction of non-guided to
guided modes, $f$, which decreases from $f = 11\%$ at the first
reflection of the ray over $f = 2.5\%$ at the second reflection to $f
< 1\%$ at further reflections. Since the average reflection length of
non-guided modes is $l_R \approx 1.5$\,mm those modes do not
contribute to the flux transmitted by fibres longer than a few
centimeters. The absorption and emission processes in fibres are
spread out over a wide band of wavelengths and the attenuation is
known to be wavelength dependent. For simplicity only monochromatic
light is assumed in the simulation and highly wavelength-dependent
effects like Rayleigh scattering are not included explicitly.

A question of practical importance for the estimation of the light
output of a particular fibre application is its transmission
function. In the meridional approximation and substituting
$\exp(-\ln{q})$ by $\exp(1-q)$ the attenuation length can be written
as
\begin{equation}
  \Lambda_m = \cos{\theta_{\it crit}}\, \left[ 1/\Lambda_{\it bulk} +
  (1-q)\sin{\theta_{\it crit}}/2\rho \right]^{-1}\ .
\end{equation}
Only for small diameter fibres ($D \sim 0.1\,$mm) are the attenuation
lengths due to imperfect reflections of the same order as the
absorption lengths. Because of the large radii of the fibres discussed
reflection losses are not relevant for the transmission function and
the attenuation length contracts to $\Lambda_m= \Lambda_{\it bulk}
\cos{\theta_{\it crit}}$. For the simulated bulk absorption length
this evaluates to $\Lambda_m= 2.8$\,m.  The transmission function
outside the meridional approximation can be found by integrating over
the normalised path length distribution, where $dN$ represents the
number of photons per path length interval $dP$, weighted by the
exponential bulk absorption factor:
\begin{equation}
  T = \frac{1}{N} \int_{P=0}^{\infty} dN/dP\, e^{-P L_F/
	\Lambda_{\it bulk}}\, dP\ .
\end{equation}
Figure~\ref{fig:absorption} shows this transmission function versus
the ratio of fibre to absorption length, $L_F/\Lambda_m$. A simple
exponential fit, $T \propto \exp\left[-L_F/\Lambda_{\it eff}\right]$,
applied to the simulated light transmissions for a varying fibre
length results in an effective attenuation length of $\Lambda_{\it
eff}= 2.4$\,m. For $L_F/\Lambda_m \ge 0.2$ this description is
sufficiently accurate to parameterise the transmission function, at
smaller values for $L_F/\Lambda_m$ the light is attenuated faster. The
difference of order 15\% to the meridional attenuation length is
attributed to the tail of the path length distribution.

Measurements of the light attenuation in fibres proves this simple
model of a single attenuation length to be wrong. A dependence of the
attenuation length with distance usually is
observed~\cite{Davis1989}. The most important cause of this effect is
the fact that the short wavelength components of the scintillation
light is dominantly absorbed. This leads to a shift of the average
wavelength in the emission spectrum towards longer wavelengths and to
an increase in the effective attenuation length.

\subsection{Trapping efficiency and transition losses in sharply curved 
	fibres}
\label{section:curvedfibres}
One of the most important practical issues in implementing optical
fibres into compact particle detector systems are macro-bending
losses. In general, some design parameters of fibre applications,
especially if the over-all size of the detector system is important,
depend crucially on the minimum permissible radius of curvature. By
using waveguide analysis transition and bending losses have been
thoroughly investigated and a loss formula in terms of the Poynting
vector can be derived~\cite{Marcuse1976,Gambling1979}. Those studies
are difficult to extend to multimode fibres since a large number of
guided modes has to be considered. When applying ray optics to curved
multimode fibres the use of a two-dimensional model is
common~\cite{Badar1989,Badar1991A,Badar1991B}. In contrast, our
simulation method follows a three-dimensional approach.

Photons are lost from a fibre core both by refraction and
tunnelling. In the simulation only refracting photons were considered.
The angle of incidence of a light ray at the tensile (outer) side of
the fibre is always smaller than at the compressed side and photons
propagate either by reflections on both sides or in the extreme
meridional case by reflections on the tensile side only. If the fibre
is curved over an arc of constant radius of curvature photons can be
refracted, and will then no longer be trapped, at the very first
reflection point on tensile side. Therefore, the trapping efficiency
for photons entering a curved section of fibre towards the tensile
side is reduced most. Figure~\ref{fig:trap-bend} quantifies the
dependence of the trapping efficiency on the azimuthal angle, $\Psi$,
between the bending plane and the photon path for a curved fibre with
a radius of curvature $R_{\it curve}=$ 2\,cm. The azimuthal angle is
orthogonal to the axial angle and $\Psi = 0\,$rad corresponds to
photons emitted towards the tensile side of the fibre.

Figure~\ref{fig:bradius} displays the explicit dependence of the
transmission function for fibres curved over circular arcs of
90$^\circ$ on the radius of curvature to fibre radius ratio for
different fibre radii, $\rho=$ 0.2, 0.6, 1.0 and 1.2\,mm. No further
light attenuation is assumed. Evidently, the number of photons which
are refracted out of a sharply curved fibre increases very rapidly
with decreasing radius of curvature. The losses are dependent only on
the curvature to fibre radius ratio, since no inherent length scale is
involved, justifying the introduction of this scaling variable. The
light loss due to bending of the fibre is about 10\% for a radius of
curvature of 65 times the fibre radius.

The use of the meridional approximation in the bending plane in place
of a three dimensional fibre is justified by the losses being
dominantly caused by meridional rays~\cite{Gloge1972,Winkler1979}.
Figure~\ref{fig:bentfibre} shows a section of a curved fibre and the
passage of a meridional ray in the bending plane with maximum axial
angle. In this model photons are guided for axial angles
\begin{equation}
  \cos\theta < \cos\theta_O= \frac{R + 2 \rho}{R + \rho} 
  \cos\theta_{\it crit}\ ,
  \label{eq:transmission}
\end{equation}
where the subscript $O$ refers to the outer fibre wall. A
transmission function can be estimated by assuming that all photons
with axial angles $\theta > \theta_O$ are refracted out of the fibre:
\begin{equation}
  T= 1 - \frac{1}{1 + R_{\it curve}/\rho}\ 
  \frac{\cos\theta_{\it crit}}{1 - \cos\theta_{\it crit}}\ .
\end{equation} 
This transmission function is shown in figure~\ref{fig:bradius} as a
dashed line and it overestimates the light losses due the larger axial
angles allowed for skew rays. A comparable theoretical calculation
using a two-dimensional slab model and a generalized Fresnel
transmission coefficient has been performed by Badar {\it et
al}~\cite{Badar1991A}. Their plot of the power contained in the fibre
core as a function of the radius of curvature (figure~5) is similar to
our results on the transmission function in the meridional
approximation. In~\cite{Winkler1979} a ray optics calculation for
curved multimode fibres involving skew rays is presented. In this
paper a discussion on the transmission function is missing. Instead, a
plot of the power remaining in a curved fibre versus distance is shown
which gives complementary information.

For photons entering a curved section of fibre the first point of
reflection on the tensile side defines the transition angle,
$\Phi_{\it trans}$, measured from the plane of entry. The angular
range of transition angles associated with each ray is called the
transition region of the fibre. For photons emitted towards the
tensile side the transition angle is related to the axial angle and
since the angular phase space density of trapped photons is highest
close to the critical angle a good estimate is $\Phi_{\it trans}=
\theta_{\it crit} - \theta_O$. For a fibre radius $\rho=$ 0.6\,mm and
radii of curvature $R_{\it curve}=$ 1, 2, and 5\,cm the above formula
leads to transition angles $\Phi_{\it trans}=$ 0.19, 0.08 and
0.03\,rad, respectively. We attribute these discrete angles to beams
emitted in the bending plane. Photons emitted from the fibre axis
towards the compressed side are not lost at this side, however, they
experience at least one reflection on the tensile side if the bending
angle exceeds the limit $\Phi_{\it limit} = \arccos\left[R_{\it
curve}/(R_{\it curve} + 2\, \rho)\right] \approx \arccos\left[1 - 2\,
\rho / R_{\it curve}\right]$. A transition in the transmission
function should occur at bending angles between $\Phi_{\it limit}/2$,
where all photons emitted towards the tensile side have experienced a
reflection, and $\Phi_{\it limit}$, where this is true for all
photons. Figure~\ref{fig:bending} shows the transmission as a function
of bending angle, $\Phi$, for a standard fibre as defined before. Once
a sharply curved fibre with a ratio $R_{\it curve}/\rho > 83$ is bent
through angles $\Phi \sim \pi/8$\,rad light losses do not increase any
further. The transition region ranges from $\sim 0.44$ to $\sim
1.06$\,rad and is indicated in the figure by arrows. At much smaller
ratios $R_{\it curve}/\rho$ the model is no longer valid to describe
this behaviour.

Experimental results on losses in curved multimode fibres along with
corresponding predictions are best known for silica fibres with core
radii $\rho \approx 50\,\mu$m.  Calculations on the basis of ray
optics for a plastic fibre with $\rho = 0.49\,$mm can be found
in~\cite{Badar1991B}. Our result on the transmission function in the
meridional approximation $T= 0.35$ at $\rho/R_{\it curve}= $20 is in
good agreement with the two-dimensional calculation. The larger value
of $T= 0.65$ predicted by the simulation is explained by the small
loss of skew rays, clearly seen in figure~\ref{fig:rlambda}. It should
be noted that the difference between finite and infinite cladding and
the appearance of oscillatory losses in the transition region has not
been investigated in the simulation.

Figure~\ref{fig:phasespace} shows contours of the angular phase space
for photons which were trapped in the straight fibre section but are
refracted out of sharply curved fibres with radii of curvature $R_{\it
curve}=$ 2 and 5\,cm. The contours demonstrate that only skew rays
from a small region close to the boundary curve are getting lost. The
smaller the radius of curvature, the larger the affected phase space
region.

\subsection{Light dispersion}
The timing resolution of scintillators are often of paramount
importance, but a pulse of light, consisting of several photons
propagating along a fibre, broadens in time. In active fibres, three
effects are responsible for the time distribution of photons reaching
the fibre exit end. Firstly the decay time of the fluorescent dopants,
usually of the order of a few nanoseconds, secondly the chromatic
dispersion in a dispersive medium, and thirdly the fact that photons
on different paths have different transit times to reach the fibre
exit end, known as inter-modal dispersion.

The chromatic dispersion is due to the spectral width,
$\Delta\lambda$, of the emitter. It is the combination of material
dispersion and waveguide dispersion. If the core refractive index is
explicitly dependent on the wavelength, $n(\lambda)$, photons of
different wavelengths have different propagation velocities along the
same path, called material dispersion. The broadening of a pulse is
given by $\Delta \tau= L_F/c_{\it core} \left( \lambda^2
d^2n/d\lambda^2 \right) \Delta\lambda/\lambda$~\cite{Ghatak1998}. The
FWHM of the emission peaks of scintillating or wavelength-shifting
fibres is approximately $40-50$\,nm. The material dispersion in the
used polymers (mostly polystyrene) is of the order of
ns$/$nm$\times$km and thus negligible for multimode fibres.

The transit time in ray optics is simply given by $\tau=
P(\theta)/c_{\it core}$, where $c_{\it core}$ is the speed of light in
the fibre core. The simulation results on the transit time are shown
in figure~\ref{fig:timing}. The full widths at half maximum (FWHM) of
the pulses in the time spectrum are presented for four different fibre
lengths. The resulting dispersion has to be compared with the time
dispersion in the meridional approximation which is simply the
difference between the shortest transit time $\tau(\theta= 0)$ and the
longest transit time $\tau(\theta= \theta_{\it crit})$: $\Delta \tau=
L_F/c_{\it core}\ (\sec{\theta_{\it crit}}-1)$, where $L_F$ is the
total axial length of the fibre. The dispersion evaluates for the
different fibre lengths to 197\,ps for 0.5\,m, 393\,ps for 1\,m,
787\,ps for 2\,m and 1181\,ps for 3\,m. Those numbers are in good
agreement with the simulation, although there are tails associated to
the propagation of skew rays. With the attenuation parameters of our
simulation the fraction of photons arriving later than $\tau(\theta=
\theta_{\it crit})$ decreases from 37.9\% for a 0.5\,m fibre to 32\%
for a 3\,m fibre due to the stronger attenuation of the skew rays in
the tail. Due to inter-modal dispersion the pulse broadening is quite
significant.

\section{Summary}

We have simulated the propagation of photons in straight and curved
optical fibres. The simulations have been used to evaluate the loss of
photons propagating in fibres curved in a circular path in one
plane. The results show that loss of photons due to the curvature of
the fibre is a simple function of radius of curvature to fibre radius
ratio and is $< 10\%$ if the ratio is $> 65$. The simulations also
show that for larger ratios this loss takes place in a transition
region ($\Phi \sim \pi/8$) during which a new distribution of photon
angles is established. Photons which survive the transition region
then propagate without further losses.

We have also used the simulation to investigate the dispersion of
transit times of photons propagating in straight fibres. For fibre
lengths between 0.5 and 3\,m we find that approximately two thirds of
the photons arrive within the spread of transit times which would be
expected from the use of the simple meridional ray approximation and
the refractive index of the fibre core. The remainder of the photons
arrive in a tail at later times due to their helical paths in the
fibre. The fraction of photons in the tail of the distribution
decreases only slowly with increasing fibre length and will depend on
the attenuation parameters of the fibre.

We find that when realistic bulk absorption and reflection losses are
included in the simulation for a straight fibre, the overall
transmission can not be described by a simple exponential function of
propagation distance because of the large spread in optical path
lengths between the most meridional and most skew rays.

We anticipate that these results on the magnitude of transition losses
will be of use for the design of particle detectors incorporating
sharply curved active fibres.

\ackn This research was supported by the UK Particle Physics and Astronomy
Research Council (PPARC).


\clearpage
\newpage


%
%


%
\begin{figure}
  \tt  \setlength{\unitlength}{\bigskipamount}
  \begin{picture}(20,23)(0,-23)
	\put(0,-1)     {Define fibre parameters: }
 	\put(0,-2)     {- Straight and bent section length, $\rho$,
			$n_{\it core}$, $n_{\it clad}$, $R_{\it curve}$}
	\put(0,-3)     {\framebox{LOOP} over photons }
	\put(1,-4)     {Generate one photon: }
	\put(1,-5)     {- Position and angle distribution according to 
			emitter type}
	\put(1,-6)     {\framebox{LOOP} over reflections at core-cladding 
			interface }
	\put(2,-7)     {Calculate photon parameters at point $\vec{r}$: }
	\put(2,-8)     {- axial and azimuthal angle, skew angle}
	\put(2,-9)     {Find next reflection point $\vec{r}_R$:
			$\longrightarrow$ \framebox{numerical methods} }
	\put(2,-10)    {Photon absorbed on path? }
	\put(2.7,-10.7){\oval(2,1)} 
	\put(2,-11)    {YES}
	\put(1.7,-10.7){\line(-1,0){1.2}}
	\put(2,-12)    {Photon reached fibre end face?}
	\put(2.7,-12.7){\oval(2,1)}
	\put(2,-13)    {YES}
	\put(1.7,-12.7){\line(-1,0){1.2}}
	\put(2,-14)    {Photon reflected at $\vec{r}_R$? }
	\put(2.4,-14.7){\oval(1.5,1)}
	\put(2,-15)    {NO}
	\put(1.65,-14.7){\line(-1,0){1.15}}
	\put(2,-16)    {Calculate propagation parameters: }
	\put(2,-17)    {- reflection length, total path length, no.\ of 
			reflections }
	\put(2,-18)    {Coordinate transformation $\vec{r}_R 
			\rightarrow \vec{r}$ }
	\put(1,-19)    {\framebox{RETURN}}
	\put(1.2,-18.2){\vector(0,1){11.95}} 
	\put(0,-20.2)  {\framebox{RETURN}}
	\put(0.5,-19.4){\vector(0,1){16.15}} 
	\put(0,-21.2)  {Calculate flux parameters: }
	\put(0,-22.2)  {- bending, absorption and reflection losses,
			trapping efficiency }
  \end{picture}
  \caption{A scheme on the main steps of the program to generate and track 
	photons in fibres of constant curvature. The equations to find the
	intersection of the light ray with the core-cladding interface
	are solved by numerical methods.}
  \label{fig:scheme}
\end{figure}
%


%
\begin{figure}[htbp]
  \begin{center}
    \subfigure[]{
	\epsfig{width= 0.47 \textwidth, file= 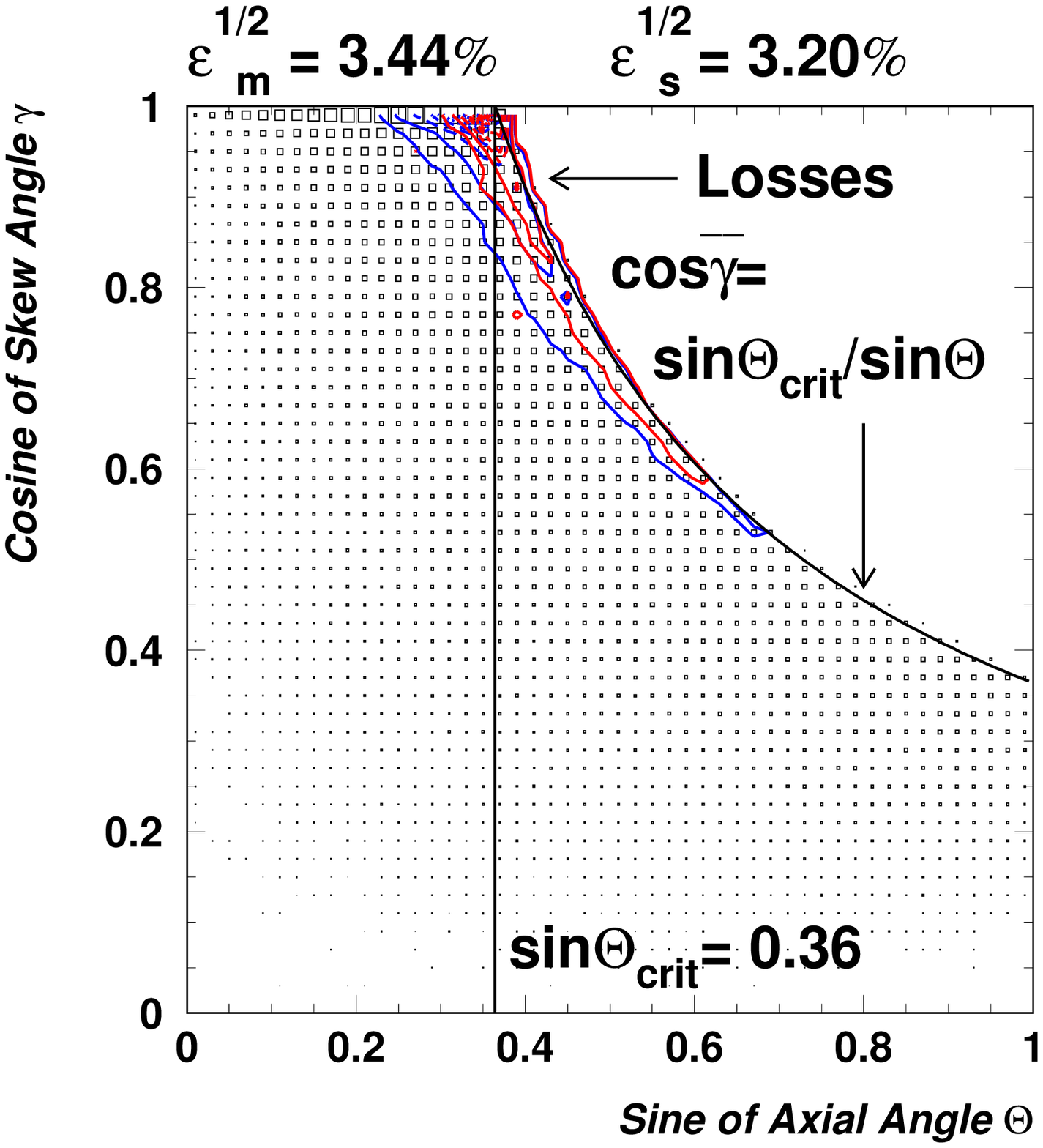} }
    \subfigure[]{
	\epsfig{width= 0.47 \textwidth, file= 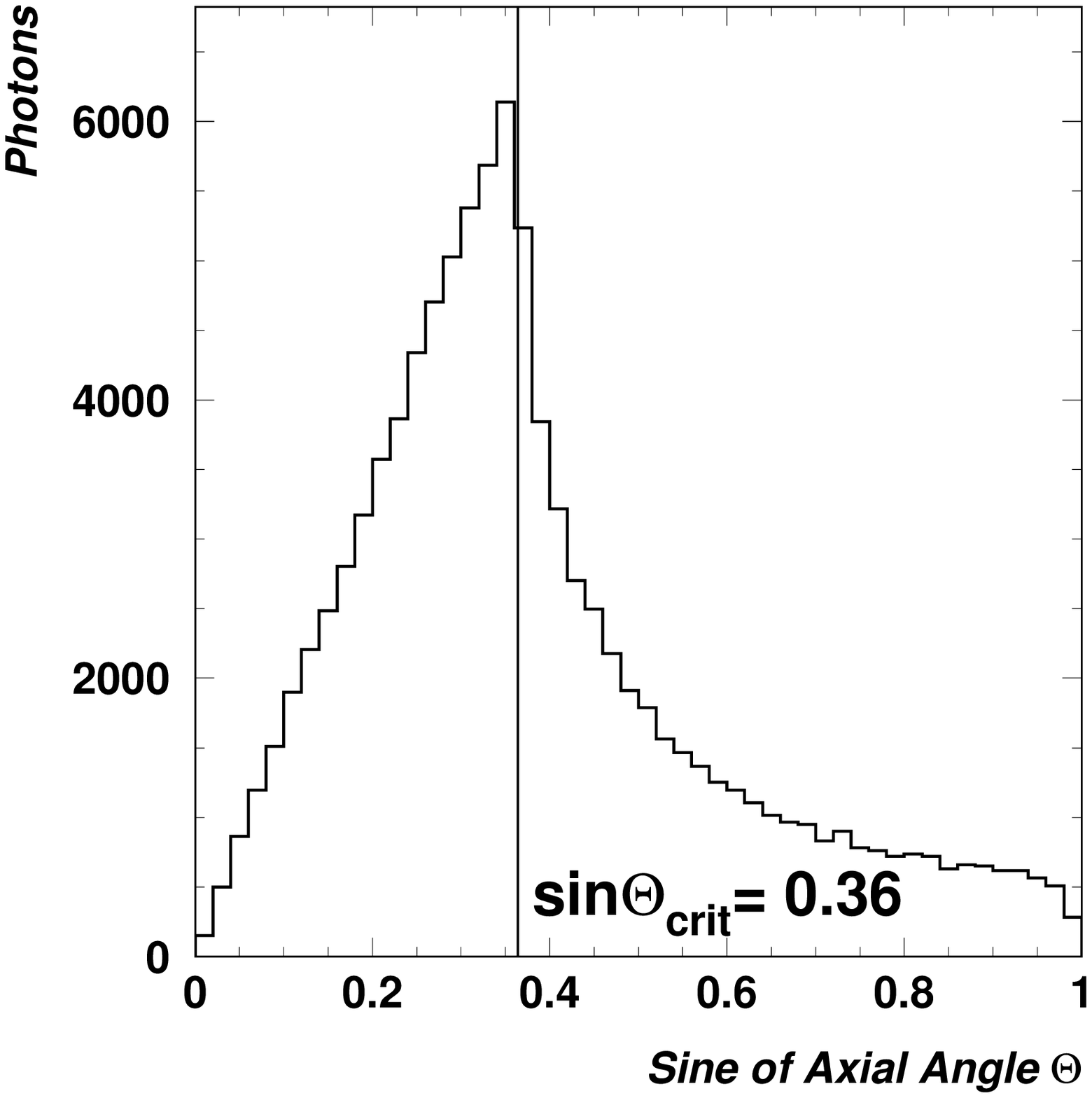} } 
    \caption{ (a) The angular phase space domain for trapped photons 
	in a fibre. To the left of the dividing line at 
	$\sin{\theta_{\it crit}}$ all skew angles are accepted. To the 
	right of the line a minimum skew angle is required by 
	the critical angle condition. The trapping efficiencies 
	are evaluated by integration over the two regions. The label
	`Losses' points to contours for photons refracted 
	out of sharply curved fibres with radii of curvature 
	$R_{\it curve}=$ 2 and 5\,cm. (b) A projection of the 
	phase space onto the $\sin\theta$-axis.}
    \label{fig:phasespace}
  \end{center}
\end{figure}
\begin{figure}[htbp]
  \begin{center}
    \subfigure[]{\label{fig:trap-r}
      \epsfig{width= 0.47 \textwidth, file= 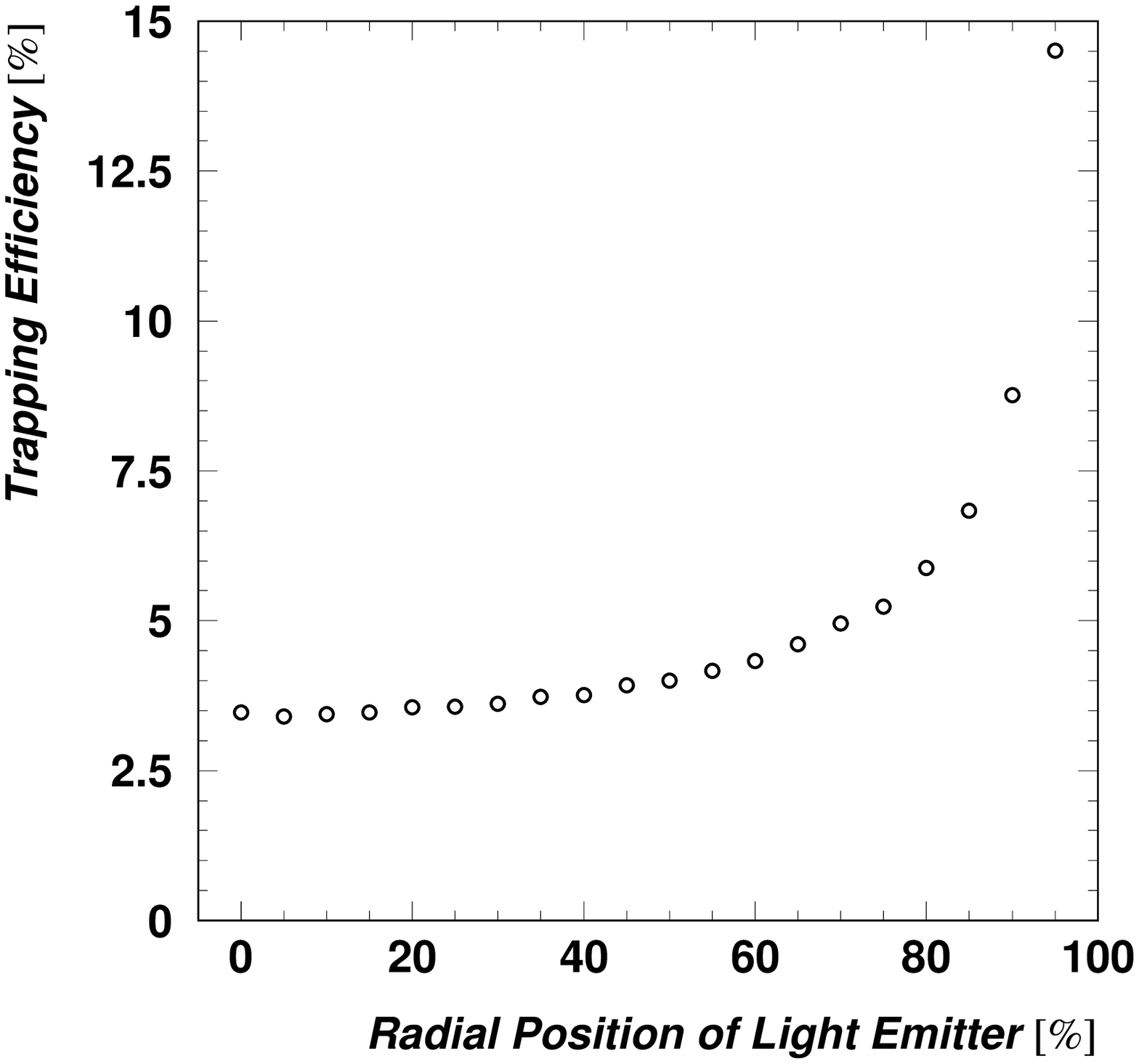} }
    \subfigure[]{\label{fig:trap-theta}
      \epsfig{width= 0.47 \textwidth, file= 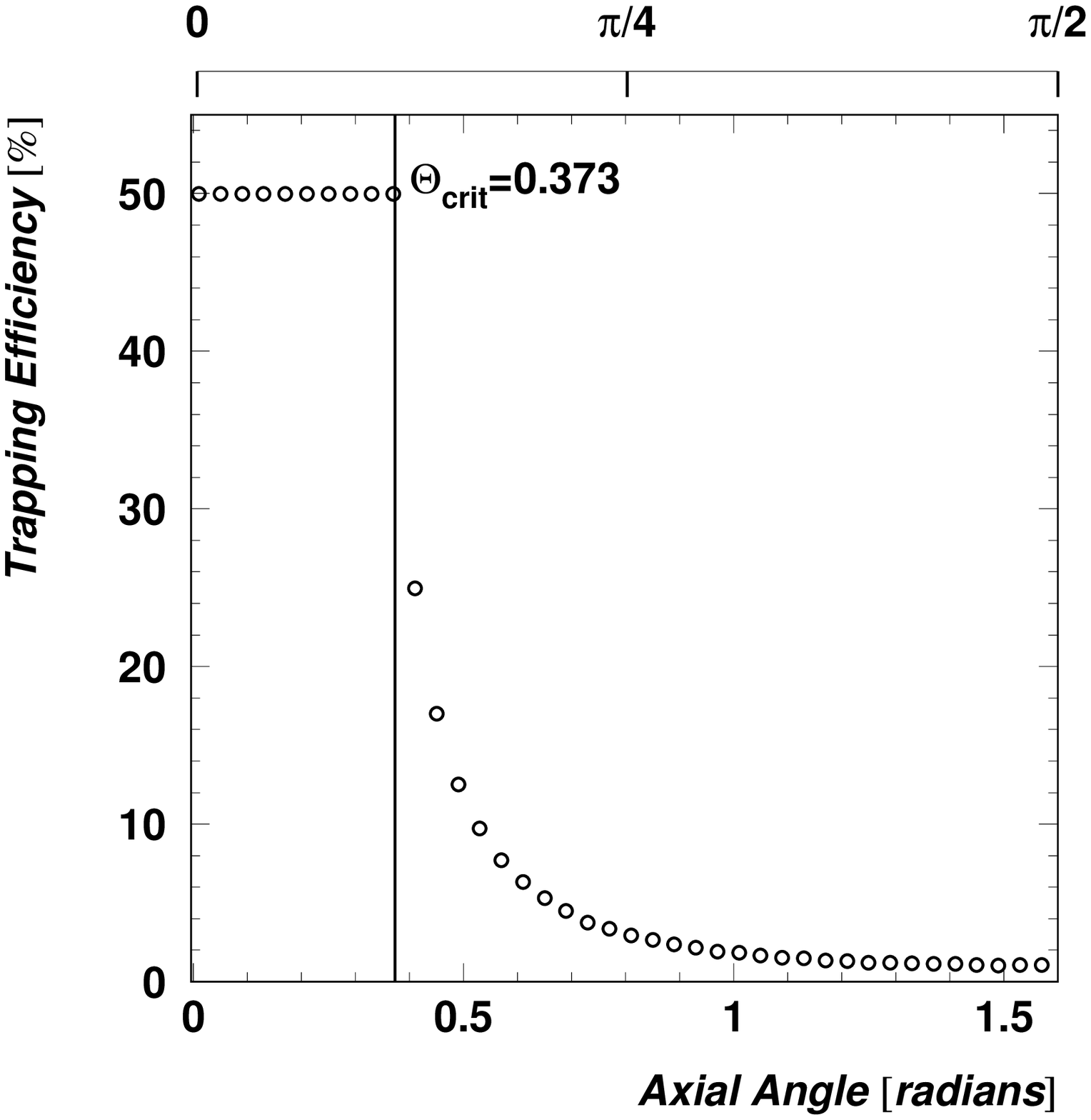} }
    \caption{ Trapping efficiency, $\epsilon^{1/2}$, for photons 
	propagating in the forward direction as a function of 
	radial position, $\hat{\rho}$, of the light emitter in 
	the fibre core (a) and of the axial angle (b). The maximum 
	axial angle allowed by the critical angle condition in 
	the meridional approximation is indicated.}
  \end{center}
\end{figure}
%


%
\begin{figure}[htbp]
  \begin{center}
    \epsfig{width= 0.5 \textwidth, file= 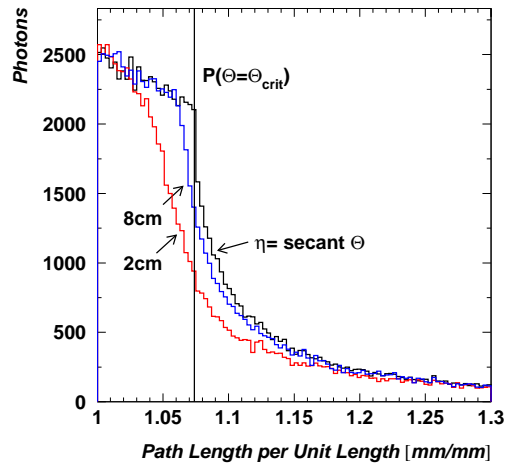}
    \caption{ The distribution of the optical path length, $P(\theta)$, 
	of trapped photons in fibres of radius $\rho=$ 0.6\,mm 
	normalised to the axial length of the fibre. The figure shows 
	$P(\theta)$ for a straight fibre and for two different radii 
	of curvature, $R_{\it curve}=$ 2 and 8\,cm. The vertical line at 
	$P(\theta_{\it crit})=$ 1.074 indicates the upper limit of 
	$P$ in the meridional approximation.}
    \label{fig:pathlength}
  \end{center}
\end{figure}
\begin{figure}[htbp]
  \begin{center}
    \epsfig{width= 0.5 \textwidth, file= 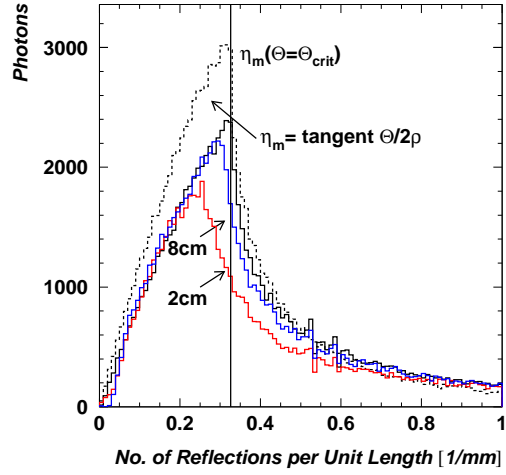}
    \caption{ The distribution of the number of reflections, 
	$\eta(\theta)$, for trapped photons in fibres of radius 
	$\rho=$ 0.6\,mm normalised to the axial length of the fibre. 
	The figure shows $\eta(\theta)$ for a straight fibre and 
	for two different radii of curvature, $R_{\it curve}=$ 2 
	and 8\,cm. The vertical line at $\eta_m (\theta_{\it crit})=$ 
	0.326 indicates its upper limit in the meridional approximation. 
	The dashed line shows the distribution of 
	$\eta_m(\theta)= \tan{\theta}/2\rho$.}
    \label{fig:reflections}
  \end{center}
\end{figure}
\begin{figure}[htbp]
  \begin{center}
    \subfigure[]{
      \epsfig{width= 0.47 \textwidth, file= 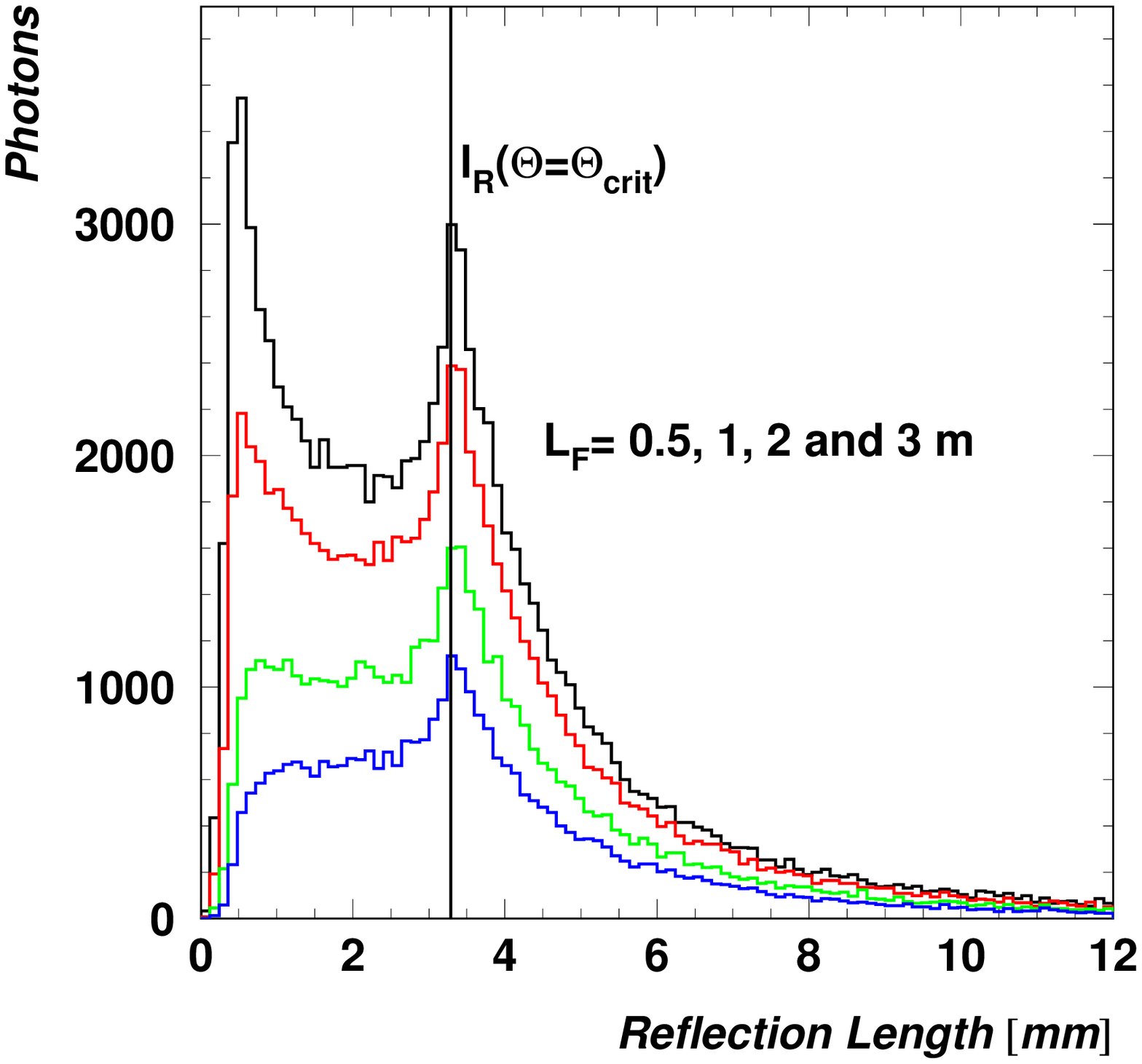} }
    \subfigure[]{
      \epsfig{width= 0.47 \textwidth, file= 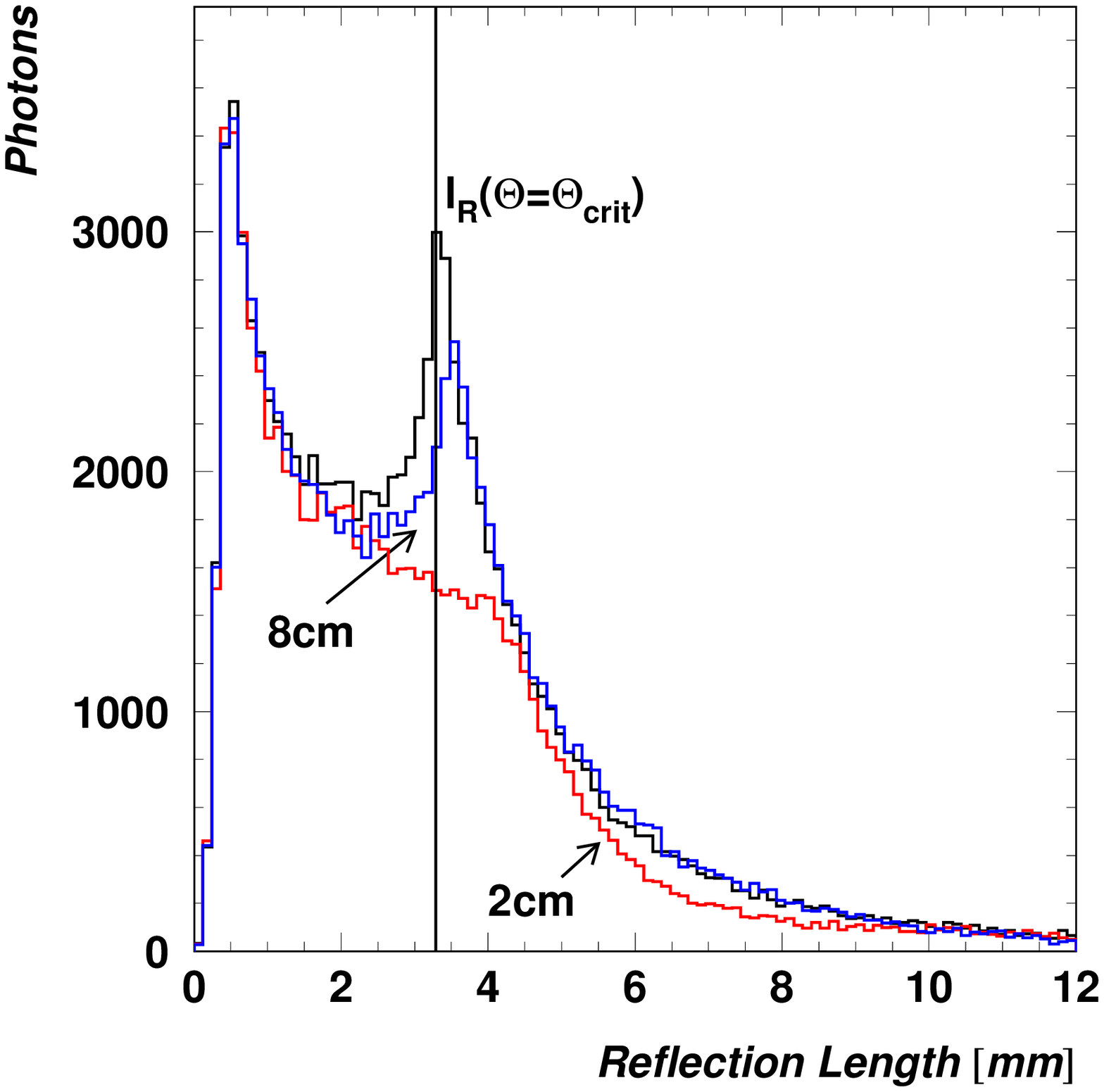} }
    \caption{ The distribution of the reflection length, 
	$l_R(\theta)$, for photons reaching the exit end of 
	fibres of radius $\rho=$ 0.6\,mm. The figure shows 
	$l_R(\theta)$ for straight fibres (a) with three 
	different fibre lengths, $L_F=$ 0.5, 1, 2 and 3\,m and 
	for curved fibres (b) of 0.5\,m length with radii of 
	curvature $R_{\it curve}=$ 2 and 8\,cm including a comparison 
	to the straight fibre. In both figures
	the vertical line at $l_R(\theta_{\it crit})=$ 
	3.29\,mm indicates the lower limit of $l_R$ in the meridional 
	approximation.}
    \label{fig:rlambda}
  \end{center}
\end{figure}
\begin{figure}[htbp]
  \begin{center}
      \epsfig{width= 0.5 \textwidth, file= 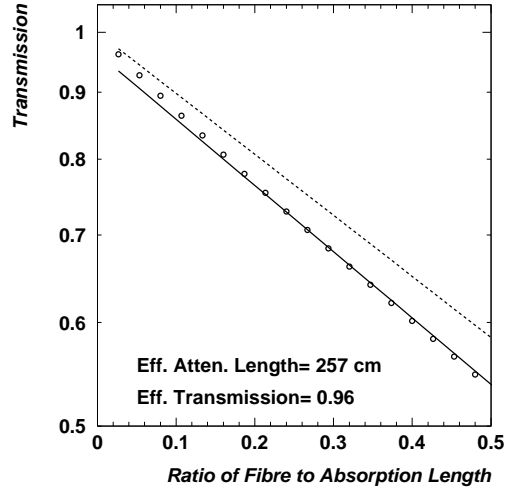}
      \caption{ Transmission function for a straight fibre 
	with a bulk light absorption length $\Lambda_{\it bulk}=$ 3\,m 
	and a reflection coefficient $q=$ 0.9999. The transmission 
	as a function of the ratio of fibre to absorption length, 
	$L_F/\Lambda_m$, is calculated from the optical
	path length distribution. A simple exponential fit results in 
	an effective attenuation length of $\Lambda_{\it eff}=$ 2.4\,m. 
	The dashed line shows the transmission function in the 
	meridional approximation with $\Lambda_m=$ 2.8\,m.}
      \label{fig:absorption}
  \end{center}
\end{figure}
%


%
\begin{figure}[htbp]
  \begin{center}
    \epsfig{width= 0.6 \textwidth, file= 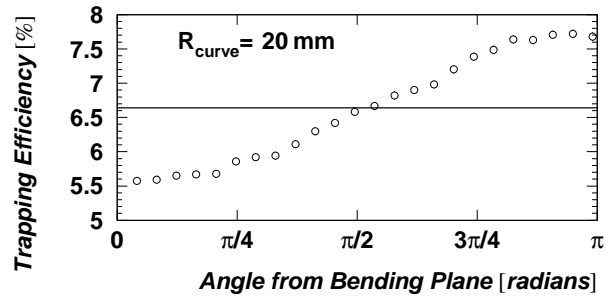}
    \caption{ Trapping efficiency, $\epsilon^{1/2}$, for photons 
	in a sharply curved fibre with radius of curvature 
	$R_{\it curve}=$ 2\,cm and fibre radius $\rho=$ 0.6\,mm. 
	The trapping efficiency is plotted as a function of the 
	azimuthal angle, $\Psi$, between the photon path and the 
	bending plane, so that $\Psi = 0\,$rad corresponds to 
	photons emitted towards the tensile side of the fibre.}
    \label{fig:trap-bend}
  \end{center}
\end{figure}
\begin{figure}[htbp]
  \begin{center}
    \epsfig{width= 0.6 \textwidth, file= 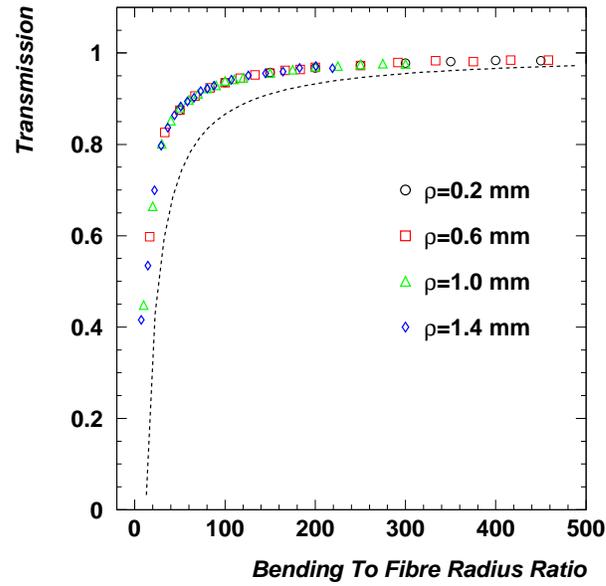}
    \caption{ The transmission function for fibres curved over a circular
	arc of 90\,$^\circ$ is plotted as a function of the radius of 
    	the curvature to fibre radius ratio for different fibre radii, 
	$\rho=$ 0.2, 0.6, 1.0 and 1.4\,mm. The dashed line is a simple
	estimate from the meridional approximation.}
    \label{fig:bradius}
  \end{center}
\end{figure}
\begin{figure}[htbp]
  \begin{center}
    \epsfig{width= 0.4 \textwidth, file= 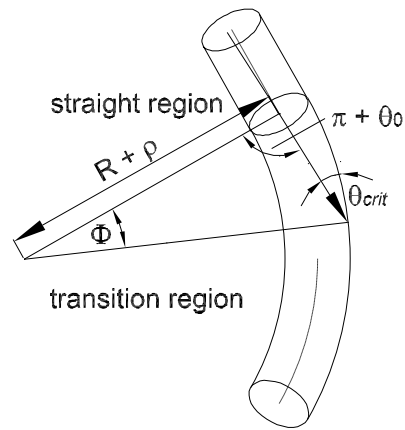}
    \caption{ Section of a curved fibre with radius $\rho$ and radius 
	of curvature $R_{\it curve}$. The passage of a meridional ray 
	in the bending plane with maximum axial angle is shown.}
    \label{fig:bentfibre}
  \end{center}
\end{figure}
\begin{figure}[htbp]
  \begin{center}
      \epsfig{width= 0.6 \textwidth, file= 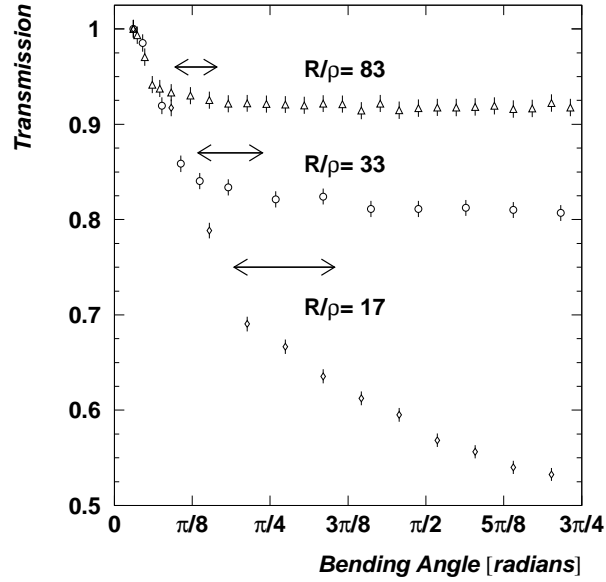}
      \caption{ Transmission function for a curved fibre 
	of radius $\rho=$ 0.6\,mm with three different radii of 
        curvature, $R_{\it curve}=$ 1, 2 and 5\,cm, corresponding to 
	the ratios $R_{\it curve}/\rho=$ 17, 33 and 83, respectively. 
	The ordinate is the fraction of photons reaching the fibre exit 
	end as a function of the bending angle, $\Phi$, and the arrows 
	indicate the transition region in the meridional approximation.}
      \label{fig:bending}
  \end{center}
\end{figure}
%


%
\begin{figure}[htbp]
  \begin{center}
      \epsfig{width= 0.6 \textwidth, file= 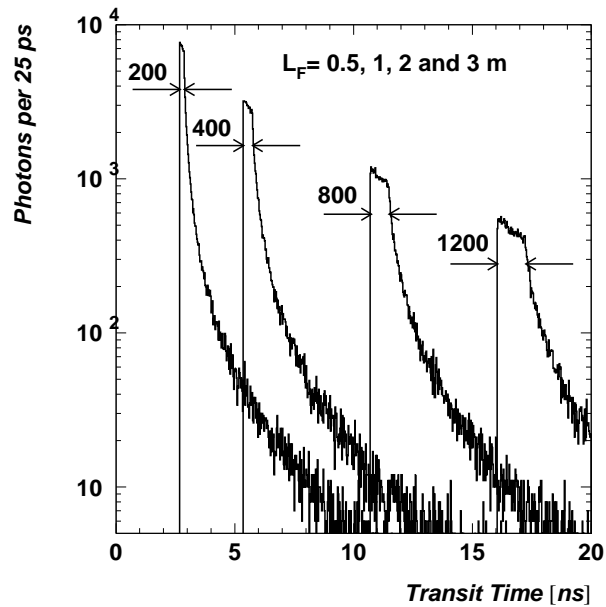}
      \caption{ The distribution of the transit time in nanoseconds 
	for photons reaching the fibre exit end. For the fibre lengths
	$L_F=$ 0.5, 1, 2 and 3\,m the pulse dispersion (FWHM) of the 
	transit time distribution is 200, 400, 775, and 1200\,ps, 
	respectively.}
      \label{fig:timing}
  \end{center}
\end{figure}

\end{document}